**Nonlinear electrostatics: steps towards a neoclassical electron model**

**D. H. Delphenich**[1]
Physics Department, Bethany College, Lindsborg, KS 67456

The equations of electrostatics are presented in pre-metric form, and it is pointed out that if the origin of the nonlinearity is the constitutive law for the medium then the differential equations themselves remain linear, while the nonlinearity is confined to an algebraic equation. These equations are solved for a general class of electric fields that include the common textbook examples, namely, fields that are adapted to a coordinate vector field. The special forms that they then take for particular electric constitutive laws of quantum origin, namely, the constitutive laws derived from the Born-Infeld and Heisenberg-Euler Lagrangians, are then discussed. Finally, the classical problem of modeling the electron is redefined in light of the established facts of quantum physics.

### 1 Introduction

After the discovery of the electron, but before the emergence of quantum theory, there were a series of attempts made at modeling the electron in terms of classical electromagnetism, notably, the work of Abraham [**1**], Lorentz [**2**], and Poincaré [**3**]. The basic properties of the electron that were taken into account were its charge, rest mass, stability, the spherical symmetry of its static field, and the fact that its static field seemed to fit the Coulomb law to a high degree of accuracy, at least as far as the existing experiments were able to distinguish.

The basic approach was to start with the field of the electron as the fundamental object of study, and then attempt to derive the charge and mass from its various properties. In particular, the charge could be obtained by integrating the flux density of the field over any closed orientable surface that enclosed the source of the field and the rest mass would presumably be due to the total self-energy of the field.

It was in the latter construction that the classical electron models ran into difficulties. Either the source charge distribution for the electron was pointlike or it was spatially extended, and most likely spherical, due to the symmetry of the field. Although a pointlike electron satisfied the requirement of stability, nevertheless, it suffered from an infinite self-energy. Whereas an extended charge distribution, such as a spherical ball or shell of a finite, but small, radius produced a finite self-energy, nonetheless, if Coulomb's law of electrostatics was still in effect at those distances, it would be difficult to explain the stability of the charge distribution under the mutual repulsion of its constituent charge elements. The resulting figure for the classical electron radius $r_c$ was obtained by equating the rest energy of the mass to the total potential energy of the field:

$$r_c = \frac{e^2}{4\pi\varepsilon_0 m_e c^2} = 2.85 \times 10^{-15} \text{ m.} \tag{1.1}$$

---
[1] E-mail: david_delphenich@yahoo.com

Eventually, quantum theory took over as the dominant approach to the structure of atomic and sub-atomic matter, and interest in the classical electron generally waned. It was generally agreed that classical Maxwellian electrodynamics seemed inapplicable to the structure of atomic nuclei and electrons, except for some aspects, such as the Coulomb form for the electrostatic field of the nucleus at the distances of the electron shells. The form that quantum electrodynamics gradually took on was an essentially phenomenological form that attempted to deduce the nature of elementary matter from results of experiments rather than postulate a basic set of field equations whose solutions would have properties that would duplicate those experimental results.

The purpose of the following discussion is to make an attempt at going beyond the Maxwellian theory by using some of the more established lessons of quantum electrodynamics. Some of these lessons must include: the wavelike nature of the electron, the existence of its anti-particle and the polarization of the vacuum, and the fact that the electron has not only mass and charge, but also intrinsic angular momentum – i.e., spin, – which also implies the existence of a magnetic dipole moment. We shall not address all of these aspects of the structure of elementary in the present work, but only concentrate on the modifications to classical electrostatics that follow from vacuum polarization and the existence of anti-matter.

The most reasonable direction for the generalization of Maxwellian theory is in the direction of nonlinear polarizable electromagnetic media, since the electric field strength at the small distances from the elementary source charge distributions must be quite large, and the existence of vacuum polarization seems to be a logical consequence of that fact.

In section 2, we present the equations of electrostatics in pre-metric form and show that the nonlinearity that is introduced by a constitutive law is purely algebraic in origin, although the differential equations for the electric field strength remain a system of linear differential equations. Hence, one can still solve them for a broad class of fields that are usually discussed in physics, namely, ones that are adapted to a coordinate vector field or 1-form. In section 3, we summarize some of the relevant issues that quantum electrodynamics and the Dirac theory of the electron introduces. In sections 4 and 5, we then specifically apply the methods of section 2 to two of the most widely-discussed quantum-corrected electromagnetic field Lagrangians, namely, the Born-Infeld and Heisenberg-Euler Lagrangians, when they are reduced to their electrostatic forms. Finally, in section 6, we attempt to redefine the classical problem of modeling the electron into a "post-quantum, neo-classical" problem, and comment on the limitations of the problem thus defined.

## 2 Nonlinear electrostatics

The very concept of a static electric field in a three-dimensional space involves a non-relativistic approximation. In the eyes of relativistic physics, one must necessarily choose an observer, in the form of a timelike congruence of curves, preferably geodesics, whose flow would then consist of isometries, in order to even speak of static solutions. However, that will not be as much of an issue in the present discussion, as much as the assumption that there is nothing time-varying about the nature of the source charge distribution, even though that seems inconsistent with the fact that the wavefunction of that distribution for an electron has a non-zero frequency $\omega_0 = m_e c^2 / \hbar = 0.89 \times 10^{21}$ rad/s associated with its non-zero rest-mass, even when viewed from the rest space of the distribution. Nevertheless, the actual field of an electron at rest, external to the source distribution, seems to have no such time-varying nature (unless perhaps that is what accounts for the well-established zero-point field of the vacuum in quantum electrodynamics). Hence, we shall essentially be considering only the time-averaged (i.e., r.m.s.) external field in the sequel.

We start with the "pre-metric" form of the Maxwellian electrostatics:

$$dE = 0, \qquad d\#\mathbf{D} = \#\rho, \qquad \mathbf{D} = \varepsilon(E). \tag{2.1}$$

These equations represent a three-dimensional reduction of the four-dimensional pre-metric form of the Maxwell equations (cf., Hehl and Obukhov [**4**] or Delphenich [**5**]):

$$dF = 0, \qquad d\#\mathfrak{h} = \#\mathbf{J}, \qquad \mathfrak{h} = \chi(F), \tag{2.2}$$

that pertain to the electromagnetic field strength 2-form $F$ on a four-dimensional manifold $M$, the electromagnetic excitation bivector field $\mathfrak{h}$, the constitutive law $\chi$ that couples them, and the electric four-current vector field $\mathbf{J}$. Although the use of the methodology of pre-metric electromagnetism is not actually essential in what follows, nevertheless, the fact that we can discuss the topic at hand without the use of a spatial metric helps to focus our attention on the spatial volume element and the constitutive law as the essential objects in the eyes of electrostatics. We shall return to the reduction of pre-metric electromagnetism to pre-metric electrostatics shortly after explaining the terminology in (2.1)

In equations (2.1), $E \in \Lambda_1(\Sigma)$ is the electric field strength one-form, which is defined on a three-dimensional spatial manifold $\Sigma$. We assume that $\Sigma$ is orientable and given a volume element $V \in \Lambda^3(\Sigma)$, whose form is a local coordinate chart $(U, x^i)$ is:

$$V = \alpha\, dx^1 \wedge dx^2 \wedge dx^3 = \frac{1}{3!} \alpha\, \varepsilon_{ijk}\, dx^i \wedge dx^j \wedge dx^k. \tag{2.3}$$

Common forms for the positive function $\alpha$ are 1, $r$, and $r^2 \sin\theta$, which correspond to Cartesian $(x, y, z)$, cylindrical $(r, \theta, z)$, and spherical $(r, \theta, \phi)$ coordinates, respectively.

The symbol $d$ refers to the exterior derivative operator, and the symbol # refers to the Poincaré duality isomorphism #: $\Lambda_k(\Sigma) \to \Lambda^{3-k}(\Sigma)$, $\mathbf{X} \mapsto i_\mathbf{X} V$ that associates $k$-vector fields with $3-k$-forms using the volume element. For a vector field $\mathbf{X}$, the local form of $\#\mathbf{X}$ is then:

$$\#\mathbf{X} = \frac{1}{2} \alpha X^i \varepsilon_{ijk}\, dx^j \wedge dx^k = \frac{1}{2} \alpha(X^1\, dx^2 \wedge dx^3 + X^2\, dx^3 \wedge dx^1 + X^3\, dx^1 \wedge dx^2). \tag{2.4}$$

Because the isomorphism # involves the use of $V$, the $3-k$-form #**X** is not invariant under all linear frame changes, but only the ones that preserve $V$. Such a form is sometimes referred to as a "twisted" form.

The symbol $\mathbf{D} \in \Lambda^1(\Sigma)$ denotes the electric displacement (or excitation) vector field, and $\varepsilon$: $\Lambda_1(\Sigma) \to \Lambda^1(\Sigma)$ is the electric constitutive law that associates $E$ with **D** (see, e.g., Landau, Lifschitz, and Pitaevskii [6] or Post [7]). Hence, the 2-form #**D** then represents the electric flux density associated with $E$. By the aforementioned reasons, the 2-form #**D** is a "twisted" 2-form.

The scalar function $\rho$ represents the electric charge density, as does the twisted 3-form #$\rho$ = $\rho V$. However, we shall only be concerned with the case in which the source of the electric field is topological in character, for which $\rho = 0$.

It is illuminating to see how this static three-dimensional formalism follows from the usual four-dimensional pre-metric formulation of electromagnetism. In that formulation, rather than dealing with a 1-form $E$ and a vector field **D** that are connected by an electrostatic constitutive law $\mathbf{D} = \varepsilon(E)$, one deals with a 2-form $F$ that represents the electromagnetic field strengths, a bivector field $\mathfrak{h}$, that represents the electromagnetic excitations and an electromagnetic constitutive law $\mathfrak{h} = \chi(F)$ that connects them. Since one generally has to take the divergence $\#^{-1}d\#\mathfrak{h}$ of $\mathfrak{h}$, it is also common to work with the twisted 2-form $H = \#\mathfrak{h}$, instead of the bivector field $\mathfrak{h}$. Here, the # isomorphism is based on a four-dimensional volume element – i.e., a 4-form, – instead of a 3-form, so the Poincaré duality isomorphism will take bivector fields to twisted 2-forms.

In order to derive electrostatics from electrodynamics, one must first define a 1+3 splitting of the tangent bundle $T(M) = L(M) \oplus \Sigma(M)$, which usually is associated with a choice of a congruence of curves – that is, an observer – whose velocity vector field $\partial_t = \partial/\partial_t$ generates the sub-bundle $L(M)$, while the spatial complement $\Sigma(M)$ must be chosen arbitrarily, in the absence of a metric. Such a choice is locally equivalent to a choice of non-zero temporal 1-form $dt$ whose annihilating subspaces are the fibers of $\Sigma(M)$. One usually chooses it so that $dt(\partial_t) = 1$, moreover.

This 1+3 splitting of $T(M)$ then induces a 3+3 splitting of both $\Lambda_2(M) = \Lambda_2^{\mathbf{H}}(M) \oplus \Lambda_2^{\mathbf{D}}(M)$ and $\Lambda^2(M) = \Lambda_E^2(M) \oplus \Lambda_B^2(M)$ into essentially electric and magnetic sub-bundles. Hence, $F$ and $\mathfrak{h}$ can be expressed in the form:

$$F = dt \wedge E - \#\mathbf{B}, \qquad \mathfrak{h} = \partial_t \wedge \mathbf{D} + \#^{-1}H, \tag{2.5}$$

in which it is important to note that the magnetic parts of $F$ and $\mathfrak{h}$ are actually the *three*-dimensional Poincaré duals of spatial vector fields and 2-forms, respectively.

Hence, in the electrostatic approximation $\mathbf{B} = 0$ and $\mathbf{H} = 0$ and we see that what we are left with is the 2-form $dt \wedge E$ and the bivector field $\partial t \wedge \mathbf{D}$, which means that they can be associated with the spatial 1-form $E$ and the spatial vector field **D**. Similarly, one sees that the electromagnetic constitutive law $\chi: \Lambda^2(M) \to \Lambda_2(M)$ reduces to an isomorphism of $\Sigma^*(M)$, viz., the spatial 1-forms, with $\Sigma(M)$. What makes this even more interesting is the fact that whereas the electromagnetic constitutive law $\chi$ only implies a *Lorentzian* metric on $T(M)$ indirectly as a result of the dispersion law that one derives from the field equations, nevertheless, the electrostatic constitutive law defines essentially a *Euclidian* spatial metric directly, by way of this isomorphism $\varepsilon: \Sigma^*(M) \to \Sigma(M)$, at least when $\varepsilon$ defines a linear and symmetric constitutive law. Specifically, one defines the metric on $\Sigma^*(M)$ by way of:

$$\varepsilon(E, E') = \varepsilon(E)(E') = \varepsilon^{ij} E_i E'_j. \tag{2.6}$$

Hence, it is unnecessary to introduce a metric in the case of electrostatics, just as it is in electrodynamics, since the constitutive law serves the same purpose.

One always assumes that $\varepsilon$ is an invertible map from any vector space $\Lambda^1_x(\Sigma)$ of 1-forms (i.e., covectors) at a given point $x \in \Sigma$ to the vector space $\Lambda_{1,x}(\Sigma)$ of vectors at that same point. When this association is linear, one can give the association the local form:

$$D^i(x) = \varepsilon^{ij}(x) E_j, \tag{2.7}$$

and when it is, moreover, isotropic, this becomes:

$$D^i(x) = \varepsilon(x) \delta^{ij} E_j. \tag{2.8}$$

Hence, one sees that the Euclidian metric $\delta = \delta^{ij} \partial_i \otimes \partial_j$ on the cotangent spaces is conformal to the metric $g = \varepsilon \delta^{ij} \partial_i \otimes \partial_j$ that is associated with an isotropic linear constitutive law. That is, the metrics are related by a relationship of the form $g = \Omega^2 \delta$, where $\Omega$ is a smooth function that is called the *conformal factor*, and equals $\sqrt{\varepsilon}$, in the present case.

As for the 1+3 decomposition of the source current vector field **J**, it takes the form $\mathbf{J} = \rho \partial_t +$ **i**, where $\rho$ is the electric potential function and **i** is a spatial electric current vector field. Its four-dimensional Poincaré dual (twisted) 3-form #**J** is then of the form $\rho V +$ #**i**, where #**i** is a twisted temporal 3-form. Hence, it has the form of $dt \wedge$ #**i**, where # now refers to the three-dimensional spatial Poincaré duality. In the electrostatic approximation, $\mathbf{i} = 0$ and the source 3-form becomes simply $\rho V$.

Returning to the electrostatic equations, we can absorb the third equation in (2.1) into the second to obtain the pair of equations:

$$dE = 0, \qquad d\#\varepsilon(E) = \rho V, \tag{2.9}$$

which then take the local component form:

$$E_{i,j} - E_{j,i} = 0, \qquad (\alpha \varepsilon(E_i))_{,i} = 0 \tag{2.10}$$

in the absence of sources.

As long one is not assuming any residual electric polarization that would make $\mathbf{D} \neq 0$ when $E = 0$, which is the case for ferroelectric media, one can model a nonlinear constitutive law locally by:

$$D^i(x) = \varepsilon^{ij}(x, E_j) E_j. \tag{2.11}$$

For such a nonlinear constitutive law, the local form of $\varepsilon$ is a 3×3 matrix whose components are functions of position and field strength and which is invertible for all possible values of these variables.

We then re-write (2.10) in the form:

$$E_{i,j} - E_{j,i} = 0, \qquad (\alpha \varepsilon^{ij} E_j)_{,i} = 0. \tag{2.12}$$

It is important to understand that equations (2.1) clearly illustrate the fact that the nonlinearity in the field equations is algebraic in nature, and not differential, since one can, of course, obtain a nonlinear differential equation by performing the differentiation in $d\#\mathbf{D}$. However, since nonlinear differential equations are generally more complicated to deal with, it is better to treat (2.1) as a set of underdetermined *linear* differential equations for the covector field $E$ and the vector field $\mathbf{D}$, together with a set of nonlinear algebraic equations that relate them.

These equations become even simpler if one considers the class of solutions for which the 1-form $E$ is adapted to the first coordinate:

$$E = E_1(x^1)\, dx^1, \qquad E_2 = E_3 = 0. \tag{2.13}$$

(This is essentially the "Gaussian pillbox" construction.) This clearly satisfies the equations $dE = 0$. This class of solutions then includes the traditional textbook examples of planar, linear, and point-like source distributions.

If we further assume that $\alpha = \alpha_1(x^1)\, \alpha_2(x^2)\, \alpha_3(x^3)$ and that the matrix $\varepsilon^{ij}$ is diagonal in the chosen coordinate system then the divergence equation in (2.10) becomes:

$$\frac{d(\alpha_1 \varepsilon^{11} E_1)}{dx^1} = 0, \tag{2.14}$$

which can be integrated to give:

$$\alpha_1 \varepsilon^{11} E_1 = C, \tag{2.15}$$

in which $C$ is the integration constant. We then rewrite this equation in the form:

$$\varepsilon^{11}(E_1)\, E_1 = \frac{C}{\alpha_1(x^1)}. \tag{2.16}$$

The problem of finding $E$ as a function of $x^1$ then reduces to the algebraic problem of solving (2.16) for $E_1(x^1)$. Although the invertibility of our constitutive law implies that such a solution will always exist, nevertheless, the actual solution might only be obtainable by numerical methods.

In order to account for the integration constant $C$, we need to address the fact that we did not include a contribution from a charge distribution in our original equations (2.1). This is because we shall choose to give the charge distribution that serves as the source of the field $\mathbf{D}$ a *topological* origin. Suppose that $S$ is a closed 2-cycle in $\Sigma$, that is, a compact orientable surface without boundary. We define the total electric flux through $S$ to be:

$$\Phi[S] = \int_S \#\mathbf{D}. \tag{2.17}$$

If $S$ bounds a 3-chain $B$ – i.e., a compact orientable 3-manifold with boundary – then one can apply Stokes's theorem to this integral (which is Gauss's law for the vector field $\mathbf{D}$) and obtain:

$$\Phi[S] = \int_B d\#\mathbf{D}. \tag{2.18}$$

If we had set $d\#\mathbf{D} = \#\rho$, where $\#\rho$ represents a charge density 3-form then we could claim that:

$$\Phi[S] = Q[B], \tag{2.19}$$

where $Q[B]$ is the total charge contained in $B$.

However, it is often the case that the field $\mathbf{D}$ is not defined at some points in $B$. For instance, this is true in the cases of a point charge, infinite line charge, and infinite surface charge. In such cases, the region of space $B$ is not actually a 3-chain; in particular, it is generally not compact. Hence, one can not apply Stokes's theorem.

Therefore, we simply *define* the charge $Q$ in the region $B$ to be $\Phi[S]$, which will still be independent of the choice of $S$, as long as $S$ always encloses the same source points, due to the homotopy invariance of the integral. One then sees that the integration constant $C$ will equal:

$$C = \frac{Q}{\int \alpha_2 \alpha_3 dx^2 \wedge dx^3}, \tag{2.20}$$

in which the domain of the integral in the denominator depends upon the choice of coordinates.

For instance, in order to obtain the classical vacuum expressions for the electrostatic field of a planar, linear, or pointlike source, resp., one chooses Cartesian, cylindrical, and spherical coordinates, resp., and lets $\varepsilon^{11}(E_1)$ be the constant $\varepsilon_0$ ($= 8.98 \times 10^{-12}$ C$^2$/N-m$^2$). One then obtains:

$$C = \frac{\sigma}{2\varepsilon_0}, \frac{\lambda}{2\pi\varepsilon_0}, \frac{q}{4\pi\varepsilon_0}, \tag{2.21}$$

respectively, where $\sigma$ is the surface charge density, $\lambda$ is the linear charge density, and $q$ is the charge of the point, resp. Note that actually we have not performed the integration over the entire two-dimensional region in the first two cases, since it is not compact, but first "retracted" it to two points in the case of a plane source (a circle in the case of a line source, resp.) and then integrated over the resulting 0-cycle (1-cycle, resp.).

With these integration constants, we then obtain the conventional expressions:

$$E_x(x) = \frac{\sigma}{2\varepsilon_0}, \quad E_r(r) = \frac{\lambda}{2\pi\varepsilon_0 r}, \quad E_r(r) = \frac{q}{4\pi\varepsilon_0 r^2}, \text{ resp.,} \tag{2.22}$$

for the cases in question. Notice that in all three cases, one has:

$$(\alpha_1 \varepsilon_0 E_1)_{,1} = 0, \tag{2.23}$$

even though the electric flux through the corresponding surfaces – viz., $x$ = const. or $r$ = const. – is $Q \neq 0$, which underscores the topological nature of the source charge.

If one wishes to give nonlinear electrostatics a Lagrangian form (cf., Plebanski [**8**] for the Lagrangian form of nonlinear electrodynamics) then one assumes that $E = d\phi$ and defines a field Lagrangian $\mathcal{L} = \mathcal{L}(x^i, E)$. The associated field equations for $E$ that one obtains by varying $\phi$ then become:

$$E = d\phi, \qquad d\#\mathbf{D} = 0, \qquad \mathbf{D} = \frac{\partial \mathcal{L}}{\partial E}. \tag{2.24}$$

If we are dealing with the linear case then:

$$\mathcal{L} = \tfrac{1}{2}\varepsilon(E, E) = \tfrac{1}{2}\varepsilon^{ij} E_i E_j, \tag{2.25}$$

and the last equation in (2.24) takes the component form:
:
$$D^i = \varepsilon^{ij} E_j. \tag{2.26}$$

One must be careful about generalizing this situation to a nonlinear Lagrangian since if we replace $\varepsilon$ in (2.25) with a field-dependent bilinear functional $\tilde{\varepsilon}(E)$, we still get the field equations (2.24), but (2.26) becomes:

$$D^i = \left( \tilde{\varepsilon}^{ij} + \tfrac{1}{2} \frac{\partial \tilde{\varepsilon}^{jk}}{\partial E_i} E_k \right) E_j = \varepsilon^{ij}(E) \, E_j. \tag{2.27}$$

Hence, it would not be correct to identify the bilinear functional $\tilde{\varepsilon}$ with the electric permittivity, which is more properly associated with:

$$\varepsilon^{ij}(E) = \tilde{\varepsilon}^{ij} + \tfrac{1}{2} \frac{\partial \tilde{\varepsilon}^{jk}}{\partial E_i} E_k. \tag{2.28}$$

However, Lagrangians of the form:

$$\mathcal{L} = \tfrac{1}{2}\varepsilon(E, E) = \tfrac{1}{2}\tilde{\varepsilon}^{ij}(E) E_i E_j. \tag{2.29}$$

will figure prominently in what follows, so we use them with this caveat.

### 3. Quantum considerations

At this point in time, most physicists would agree that when dealing with Nature at the atomic to subatomic scale the results of quantum physics are more definitive than those of classical electromagnetism. Hence, we shall summarize some of the salient facts regarding the nature of the electron as it is described by quantum mechanics and quantum electrodynamics.

One of the most fundamental aspects of the Dirac theory of the electron was the notion that any particle that was represented by a Dirac spinor – i.e., any fermion – would be paired with an anti-particle that had the same mass and spin, but an opposite charge. Moreover, the interaction of any particle with its anti-particle might produce a photon in place of the particle/anti-particle pair; conversely, any sufficiently high energy photon could split into a particle/anti-particle pair in the presence of an external electric or magnetic field of high enough strength. In the transition region between having too little energy to result in pair production and having more than enough, one would expect to find vacuum polarization, which one might model by making $\varepsilon_0$ and $\mu_0$ depend upon **E** and **B**, or at least their magnitudes, unless one also intends that the symmetry of the vacuum as an electromagnetic medium is broken by the formation of particle/anti-particle pairs, which is conceivable. Most of the best-established consequences of quantum electrodynamics are traceable to precisely the existence of such a vacuum polarization process.

For instance, one can make quantum electrodynamical corrections to the Coulomb potential that are based on vacuum polarization, since presumably as one approaches an elementary charge distribution the electric field strength of that distribution approaches the critical value for electron-positron pair production. However, the applicability of quantum electrodynamics then becomes limited by the possibility that at a high enough field strength one might produce pion/anti-pion pairs, which are strongly interacting particles and therefore no longer best treated by quantum electrodynamics, but quantum chromodynamics. The one-loop corrected version of the Coulomb potential takes the form (cf., Berestetskii, et al. [**9**], or Greiner, et al. [**10**]):

$$\phi(r) = \frac{Q}{4\pi\varepsilon_0 r}\left[1 + \frac{2\alpha}{3\pi}\left(\ln\frac{\lambda_c}{r} - C - \frac{5}{6}\right)\right] \quad (3.1)$$

when $r \ll \lambda_c$ and:

$$\phi(r) = \frac{Q}{4\pi\varepsilon_0 r}\left[1 + \frac{\alpha}{4\sqrt{\pi}}\frac{e^{-2r/\lambda_c}}{\sqrt{r/\lambda_c}}\right] \quad (3.2)$$

when $r \gg \lambda_c$. Here, the symbol $\alpha$ represents the fine structure constant 1/137 for electromagnetism, which serves as the electromagnetic coupling constant, and the distance $\lambda_c$ is the Compton wavelength for the electron.

In particular, one notes that in both cases the quantum correction adds to the Coulomb potential, so $\phi(r)$ still diverges as $r$ goes to zero. One also notes that the two asymptotic expressions do not agree formally at $r = \lambda_c$, although numerically it is the difference between the bracketed term taking the value 1.0054 for the small-$r$ expression versus 0.998 for the large-$r$ expression.

One of the more celebrated consequences of vacuum polarization is the explanation for the anomalous magnetic moment of the electron, a fact that was established experimentally by the Lamb shift of the atomic spectrum. According to relativistic wave mechanics alone – viz., the Dirac equation – the electron should have a magnetic moment that is given by the Bohr magneton: $\mu_B = e\hbar / 2m_e c = 3.09 \times 10^{-32}$ J/T, including the relativistic contribution of the factor ½, which arises as a result of Thomas precession. Due to the effect of vacuum polarization on the electron form factors, this figure gets corrected to:

$$\mu = \mu_B \left( 1 + \frac{\alpha}{2\pi} - 0.328 \frac{\alpha^2}{\pi^2} \right) = \mu_B (1 + 0.00116 - 0.00000177) , \qquad (3.3)$$

when one includes all diagrams to quadratic order. However, one can see that the successive terms in the sum are each separated by about three orders of magnitude.

## 4 Born-Infeld electrostatics

In 1934, Born and Infeld [**11**] (see also Born [**12**]) attempted to remedy the problem of the infinite self-energy of a pointlike electron by assuming that as a result of vacuum polarization it was not physically possible for the electric field strength to exceed a maximum value $E_c$. The customary way of obtaining this value is to compute the Coulomb field strength at a distance from the pointlike electron that equals the classical electron radius $r_c$, which is the radius that makes the self-energy of the field equal to the rest mass. One then obtains:

$$E_c = \frac{e}{4\pi\varepsilon_0 r_c^2} = 1.78 \times 10^{20} \text{ V/m}. \qquad (4.1)$$

Of course, if one is assuming a pointlike electron then it would seem somewhat irrelevant to use a radius that is associated with an extended one, but the use of this value seems to have a long tradition.

The Born-Infeld theory starts with the field Lagrangian[2]:

$$\mathcal{L}_{BE} = \alpha \left( 1 - \sqrt{1 - \frac{\mathcal{F}}{E_c^2} - \frac{\mathcal{G}^2}{E_c^4}} \right). \qquad (4.2)$$

This time, the symbol $\alpha$ refers to the function that appears in the volume element (2.3), whose form depends upon the choice of coordinate system. The symbols $\mathcal{F}$ and $\mathcal{G}$ refer to the Lorentz-invariant scalars that one obtains from the electromagnetic field strength 2-form $F$:

$$\mathcal{F} = \chi(F, F) = \tfrac{1}{2} F_{\mu\nu} H^{\mu\nu}, \qquad \mathcal{G} = V(F, F) = \tfrac{1}{2} F_{\mu\nu} {}^*F^{\mu\nu}. \qquad (4.3)$$

---

[2] We suppress the explicit mention of the leading constant scalar multiplier in the Lagrangian, since it is only necessary in order to give the Lagrangian the units of an energy density, but does not appear in the field equations.

In the second expression, we are letting $*F^{\mu\nu} = \tfrac{1}{2} \varepsilon^{\mu\nu\kappa\lambda} F_{\kappa\lambda}$ denote the components of the bivector field $\#F$ that is Poincaré dual to the 2-form $F$. We are also generalizing the electrostatic constitutive law to an electromagnetic constitutive law, i.e., a one-to-one correspondence $\chi$: $\Lambda_2(M) \to \Lambda^2(M)$, $F \mapsto \chi(F) = H$ between 2-forms and bivector fields.

In order for this field Lagrangian to produce field equations that reduce to Maxwell's equations in the limit of small field strengths, one must assume that the 2-form $F$ is exact; that is:

$$F = dA \qquad (4.4)$$

for some potential 1-form $A$. The other field equation is then obtained by varying the field $A$, although we shall pass on to the static electric case.

When we restrict ourselves to the static electric case, in which $F = dt \wedge E$, we find that the invariants take the form:

$$\mathcal{F} = D^i E_i = \varepsilon^{ij} E_i E_j, \qquad \mathcal{G} = 0. \qquad (4.5)$$

The condition (4.4) becomes $E = d\phi$, as above, and if we vary the remaining Lagrangian, which we abbreviate to:

$$\mathcal{L}_{\text{BE}} = \alpha \sqrt{1 - \frac{\varepsilon(E,E)}{E_c^2}}, \qquad (4.6)$$

with respect to $\phi$ then the field equation becomes:

$$0 = d \, \# \left( \frac{\partial \mathcal{L}_{\text{BE}}}{\partial E} \right), \qquad (4.7)$$

which then takes the form:

$$0 = d \left[ \alpha \left( 1 - \frac{\varepsilon(E,E)}{E_c^2} \right)^{-1/2} \# \mathbf{D} \right]. \qquad (4.8)$$

We can give this the same form as our nonlinear electrostatic equation (the second equation in (2.5)) if we introduce the rescaled electric permittivity matrix:

$$\bar{\varepsilon}^{ij} = \left( 1 - \frac{\varepsilon(E,E)}{E_c^2} \right)^{-1/2} \varepsilon^{ij}. \qquad (4.9)$$

One notes that the form of this equation suggests that electric permittivity behaves as the field strength approaches its limiting values in the same way that mass behaves as its speed approaches the speed of light. However, this is not a mysterious coincidence, but a predictable consequence of the geometrical form of the Lagrangian that Born and Infeld originally chose before they settled on the form given here.

Equation (4.8) then takes the local component form:

$$(\alpha \bar{\varepsilon}^{ij} E_j)_{,i} = 0 .\tag{4.9}$$

If we duplicate the solution of this equation that we gave in the previous section in the case of spherical coordinates and an isotropic electric permittivity matrix of the form $\varepsilon_0 \delta^{ij}$ then we arrive at the algebraic equation:

$$\frac{E_r(r)}{\sqrt{1 - \frac{E_r^2}{E_c^2}}} = \frac{e}{4\pi\varepsilon_0 r^2} \equiv E_{\text{Coul}}(r),\tag{4.10}$$

which can be solved to give:

$$E_r(r) = \left(1 + \frac{E_{\text{Coul}}^2(r)}{E_c^2}\right)^{-1/2} E_{\text{Coul}}(r) ,\tag{4.11}$$

which can also be written in the form:

$$E_r(r) = \left(1 + \frac{E_c^2}{E_{\text{Coul}}^2(r)}\right)^{-1/2} E_c .\tag{4.12}$$

In the form (4.11), one can see how the Born-Infeld field strength $E_r(r)$ converges to the Coulomb field for small field strengths. In the form (4.12), one can see how $E_r(r)$ approaches the limiting value of $E_c$ as $r$ goes to 0. For the record, when $r$ equals the classical electron radius $E_r(r)$ has the value $E_c/\sqrt{2}$, although $r_c$ seems to play no role in this model beyond that of defining the value of $E_c$.

## 5. Heisenberg-Euler electrostatics

One of the early attempts to deduce phenomenological physical consequences from the Dirac theory of the electron was made by Heisenberg and Euler [**13**] in 1936. What they accomplished was to derive what would now be called a one-loop effective Lagrangian for a quantized (i.e., operator-valued) spinor field, namely, the field of an electron/positron pair, that is coupled to a constant – or at least slowly time-varying – electromagnetic field that is regarded as an external field *F*, and therefore unquantized. In effect, one then integrates out the higher-energy modes that the electron/positron pair contributes and obtains a correction to the Lagrangian for the external field. The correction $\delta\mathcal{L}$, up to first order in the coupling constant $\alpha$, which takes the form $e^2/4\pi\varepsilon_0 \hbar c$ in SI units, to the Lagrangian $\mathcal{L}$ of the external electromagnetic field that was due to the polarization of the vacuum as its field strength approached the critical value $E_c$ they obtained was:

$$\delta\mathcal{L} = \frac{\alpha}{2\pi}\int_0^\infty d\eta \frac{e^{-\eta}}{\eta^3}\left\{(E_c^2 - \frac{\eta^2}{3}\mathcal{F}) - i\eta^2\mathcal{G}\frac{\cos\left(\frac{\eta}{E_c}\sqrt{\mathcal{F}-i\mathcal{G}}\right) + c.c.}{\cos\left(\frac{\eta}{E_c}\sqrt{\mathcal{F}-i\mathcal{G}}\right) - c.c.}\right\}. \tag{5.1}$$

In this equation, $\mathcal{F}$ and $\mathcal{G}$ are the two Lorentz invariants that one can form from $F$ that we mentioned in section 3. We have also restricted the electromagnetic constitutive law of the vacuum to the classical static, homogeneous, isotropic case. In particular, this means that the classical electromagnetic field Lagrangian is:

$$\mathcal{L} = \tfrac{1}{2}\mathcal{F}. \tag{5.2}$$

When the field strengths are less than $E_c$ the Lagrangian (5.1) can be expanded in a power series, which is, to sixth order:

$$\delta\mathcal{L} = \xi\left\{\frac{1}{2}(\mathcal{F}^2 + 7\mathcal{G}^2) + \frac{1}{7E_c^2}(13\mathcal{G}^2\mathcal{F} + 2\mathcal{F}^3)\right\}, \tag{5.3}$$

into which we have introduced the abbreviation $\xi = \alpha/180\pi E_c^2 = 10^{-38}$ m$^2$/V$^2$.

Since our immediate concern is the static electric field, which makes $\mathcal{F} = \varepsilon_0 E^2$ and $\mathcal{G} = 0$, we can represent the total sixth-order one-loop corrected electric field Lagrangian as:

$$\mathcal{L} + \delta\mathcal{L} = \tfrac{1}{2}\varepsilon_0\left[1 + \frac{\alpha}{360\pi}\varepsilon_0\hat{E}^2 + \frac{\alpha}{630\pi}\varepsilon_0^2\hat{E}^4\right]E^2. \tag{5.4}$$

into which we have introduced the notation $\hat{E} = E/E_c$ for the rescaled electric field strength 1-form, or its magnitude.

If we introduce the further notation:

$$\tilde{\varepsilon}(E) = \varepsilon_0\left[1 + \frac{\alpha}{360\pi}\varepsilon_0\hat{E}^2 + \frac{\alpha}{630\pi}\varepsilon_0^2\hat{E}^4\right], \tag{5.5}$$

then we see that this field Lagrangian takes the elementary nonlinear electrostatic form (2.27). Once again, we recall that $\tilde{\varepsilon}(E)$ is not the electric permittivity, but:

$$\varepsilon(E) = \frac{\partial(\mathcal{L}+\delta\mathcal{L})}{\partial E} = \varepsilon_0\left[1 + \frac{\alpha}{180\pi}\varepsilon_0\hat{E}^2 + \frac{\alpha}{210\pi}\varepsilon_0^2\hat{E}^4\right] = \varepsilon_0 + \delta\varepsilon_0(\hat{E}). \tag{5.8}$$

We then deduce that the electric polarization vector field $\mathbf{P} = \mathbf{D} - 4\pi\varepsilon_0\mathbf{E}$ associated with $E$ has the components:

$$P^i = \frac{\partial(\delta\mathcal{L})}{\partial E_i} = \frac{\alpha}{180\pi}\varepsilon_0^2 \hat{E}^2 \left[1 + \frac{6}{7}\varepsilon_0 \hat{E}^2\right] E_i, \tag{5.9}$$

which makes the electric susceptibility $\chi = \varepsilon - 4\pi\varepsilon_0$ of the Heisenberg-Euler vacuum take the form:

$$\chi(E) = \frac{\alpha}{180\pi}\varepsilon_0^2 \hat{E}^2 \left[1 + \frac{6}{7}\varepsilon_0 \hat{E}^2\right]. \tag{5.10}$$

It is important to point out if one derives the magnetization vector field $\mathbf{M}$ from the general Lagrangian (5.3) then one obtains:

$$M^i = -\frac{\partial(\delta\mathcal{L})}{\partial M_i} = \gamma(E, B)\, E^i + \zeta(E, B)\, B^i, \tag{5.11}$$

in which:

$$\gamma(E, B) = -\frac{\alpha}{90\pi}\hat{\mathbf{E}}\cdot\hat{\mathbf{B}}\left\{14 + \frac{52}{7}(\varepsilon_0 \hat{E}^2 - \frac{1}{\mu_0}\hat{B}^2)\right\}, \tag{5.12a}$$

$$\zeta(E, B) = \frac{\alpha}{45\pi\mu_0}\left[\varepsilon_0 \hat{E}^2 - \frac{1}{\mu_0}\hat{B}^2 + \frac{26}{7}(\hat{\mathbf{E}}\cdot\hat{\mathbf{B}})^2 + \frac{3}{7}(\varepsilon_0 \hat{E}^2 - \frac{1}{\mu_0}\hat{B}^2)^2\right]. \tag{5.12b}$$

Although $\mathbf{M}$ naturally vanishes in the absence of a magnetic field, one does still have a non-vanishing magnetic susceptibility of:

$$\zeta(E) = \frac{\alpha\varepsilon_0 \hat{E}^2}{45\pi\mu_0}\left[1 + \frac{3}{7}\varepsilon_0 \hat{E}^2\right]. \tag{5.13}$$

This is consistent with the established fact of quantum physics that even a static electron has a non-vanishing magnetic dipole moment, which is, of course, explained by its non-vanishing spin in traditional quantum physics.

We can once again set up the problem of finding static electric fields that are adapted to some coordinate vector field as in the previous sections, and our nonlinear algebraic equation for $E$ now takes the form:

$$\varepsilon(E)\, E = \varepsilon_0 \left[1 + \frac{\alpha}{180\pi}\varepsilon_0 \hat{E}^2 + \frac{\alpha}{210\pi}\varepsilon_0^2 \hat{E}^4\right] E = \frac{C}{\alpha_1(x^1)}. \tag{5.14}$$

which can be put into the form:

$$\frac{\alpha}{210\pi}\varepsilon_0^2 \hat{E}^4 E + \frac{\alpha}{180\pi}\varepsilon_0 \hat{E}^2 E + E - \frac{C}{\varepsilon_0 \alpha_1(x^1)} = 0. \tag{5.15}$$

Since this equation is quintic in $E$, the only hope for finding explicit solutions would have to be numerical or perturbative. One could also restrict the approximate form (5.4) of the Heisenberg-Euler Lagrangian to fourth order and then obtain a cubic equation in place of (5.14), which could be solved explicitly, at least in principle, although we shall not do so here.

### 6. Discussion

Let us first summarize the character of the various solutions to the spherically-symmetric electrostatic equations that were obtained by various choices of constitutive law. All of the constitutive laws that we discussed above are isotropic, and therefore take the form $\varepsilon^{ij} = \varepsilon(r, E)\delta^{ij}$ for an appropriate frame, where $E$ refers to the magnitude of the field strength. As a result, in all cases, except possibly the Heisenberg-Euler case, the resulting electric field $\mathbf{E}(r)$ took the form of $f(r, E)\mathbf{E}_{\text{Coul}}(r)$ for some appropriate function $f(r, E)$. Only the Born-Infeld solution seems to be finite at $r = 0$, although the analytical character of the Heisenberg-Euler solution must be obtained indirectly by further study.

The simplest constitutive law is $\varepsilon(r, E) = \varepsilon_0$, which is then homogeneous, isotropic, and linear. It gives the usual Coulomb solution, which is undefined at $r = 0$, since it becomes infinite as $r$ approaches 0.

The 1-loop quantum correction to the Coulomb law replaces $\varepsilon_0$ with a more general $\varepsilon(r)$, which then makes the constitutive linear and isotropic, but inhomogeneous. It does not, however, make the resulting $\mathbf{E}(r)$ finite at $r = 0$.

The Born-Infeld constitutive takes the form $\varepsilon(E)$, so it is homogeneous and isotropic, but nonlinear. It was specifically intended to make the $\mathbf{E}(r)$ field converge to a finite limiting value at $r = 0$.

The Heisenberg-Euler constitutive law is similar to the Born-Infeld law, at least when one expands Born-Infeld in a power series, so it is also homogeneous and isotropic, but nonlinear. However, since the field $\mathbf{E}(r)$ can be derived only by numerical or perturbative means, the question of how it behaves as $r$ approaches zero requires deeper analytical study.

Prior to the onset of quantum field theory the most fundamental problems in any field theory were boundary-value problems, in the static case, as well as initial-value problems, in the dynamic case. The nature of the classical electron model was that the field of an electron at rest should be the unique solution to a spherical boundary-value problem in the field equations of electrostatics that vanished at infinity, although there was some dispute over how to define the field at the source. Because this dispute could not be resolved within the context of linear Maxwellian electrostatics, and the experiments regarding the nature of atomic structure suggested that quite of bit more of Maxwell's theory broke down at the atomic level, interest in finding such a classical model eventually waned.

Nonetheless, there is some value in defining a difficult problem, even while knowing that its solution might not be forthcoming, if only to stimulate and focus discussion of the matter that is being addressed. Hence, we shall define the "neo-classical" electron problem to be the problem of modeling an electron – or, more generally, any charged "irreducible" particle – as a field that is a solution to a boundary-value problem in nonlinear electrostatics that agrees with the currently-accepted facts regarding the particle. One then identifies different levels of detail regarding the problem: the macroscopic level, the atomic level, and the subatomic level.

The classical electron of Abraham, Lorentz, and Poincaré was essentially valid at the macroscopic level and broke down at the atomic level. The basic properties of the electron that it attempted to account for were its charge, rest mass, stability, the spherical symmetry of its static field, and its asymptotic agreement with the Coulomb law at scales larger than atomic.

In order to proceed into the atomic level, one must keep in mind the lessons of quantum theory, which expand this list to: the wavelike nature of the electron, the existence of its anti-particle and the polarization of the vacuum, and the fact that the electron has not only mass and charge, but also intrinsic angular momentum – i.e., spin, – which also implies the existence of a magnetic dipole moment. Furthermore, one must account for the fact that this magnetic moment itself takes on an "anomalous" contribution from vacuum polarization. Admittedly, the Born-Infeld and Heisenberg-Euler models did not address the wavelike nature of the electron or its spin, at least directly, so that seems to be the most immediate direction of extension for the model.

In order to proceed to the subatomic level, one must address not only the contributions of quantum electrodynamics, but also those of quantum chromodynamics, since one will be forced to deal with the fact that the charge distributions of protons and atomic nuclei are extended, but stable to varying degrees. Although this seems to complicate matters beyond reason, perhaps by the time one has successfully achieved the extension of the classical electron to a neo-classical electron that is valid at the atomic level the nature of the next extension will seem more straightforward.